# Implementation of an alternative method for assessing competing risks: restricted mean time lost [§]

Hongji Wu[1], Hao Yuan[1], Zijing Yang[1], Yawen Hou[2], Zheng Chen[1*]

1: Department of Biostatistics, Southern Medical University, Guangzhou, China

2: Department of Statistics, Jinan University, Guangzhou, China

* Corresponding author: Zheng Chen

Dec. 2020

**ABSTRACT**

In clinical and epidemiological studies, hazard ratios are often applied to compare treatment effects between two groups for survival data. For competing risks data, the corresponding quantities of interest are cause-specific hazard ratios (cHRs) and subdistribution hazard ratios (sHRs). However, they both have some limitations related to model assumptions and clinical interpretation. Therefore, we recommend restricted mean time lost (RMTL) as an alternative that is easy to interpret in a competing risks framework. Based on the difference in restricted mean time lost (RMTLd), we propose a new estimator, hypothetical test and sample size formula. The simulation results show that the estimation of the RMTLd is accurate and that the RMTLd test has robust statistical performance (both type I error and power). The results of three example analyses also verify the performance of the RMTLd test. From the perspectives of clinical interpretation, application conditions and statistical performance, we recommend that the RMTLd be reported with the HR in the analysis of competing risks data and that the RMTLd even be regarded as the primary outcome when the proportional hazard assumption fails. The R code (crRMTL) is publicly available from Github (https://github.com/chenzgz/crRMTL.1).

**Keywords:** survival analysis, competing risks, hazard ratio, restricted mean time lost,

sample size, hypothesis test

Clinical trials of treatments and preventative measures for coronavirus disease 2019 (COVID-19) have received global attention. In published and ongoing randomized trials for COVID-19 treatments, the time-to-event endpoint of interest, such as the time to clinical improvement (or recovery), has been the most commonly used primary outcome (1). The corresponding method used has been the Kaplan-Meier method, and the effect size has been the hazard ratio (HR). However, patients may die of COVID-19 before improvement (or recovery), so competing risks problems occur (2); that is, the occurrence of the event of interest (improvement or recovery) may be precluded by a competing event (death). At this time, the commonly applied single-event survival analysis techniques may lead to biased results, with subjects who experience a competing event being censored (3, 4). Therefore, competing risks analysis should be applied in such situations.

There are two widely used approaches to competing risks analysis based on hazards (5): one is based on a cause-specific hazard function (CSH), which refers to the instantaneous rate of occurrence of a specific event among the individuals who are still event-free; its corresponding statistical test is the log-rank test, and the statistical measure, i.e., the cause-specific hazard ratio (cHR), can be estimated through a cause-specific Cox regression model. The other approach is the subdistribution hazard function (SDH), which refers to the instantaneous rate of the event of interest in subjects who have not yet experienced the given event. The statistical test is the Gray test, and the estimated effect of one group relative to another, i.e., the subdistribution hazard ratio (sHR), can be calculated by the Fine-Gray model. Meanwhile, the descriptive statistics for clinical or epidemiological interests in this approach are described by the cumulative incidence function (CIF), the probability of one event of interest occurring by a particular time in the presence of other events, which reflects the risk of the cause of interest without ignoring the presence of other competing events.

In the clinical analysis of competing risks data, the estimations and statistical tests based on cHR and sHR still have some limitations: 1) The hazard ratio (both the cHR and the sHR) should be described as a relative rate, not as a relative risk (6). Without the assumption of independence of competing events, the cHR cannot be linked to the comparison of CIFs for an event between two groups (7), i.e., cHR > 1 does not necessarily imply $CIF_1 > CIF_0$; that is, even if the hazard due to a main cause in a control group is always higher than that in a treatment group, the risk of the main cause in the control group is not necessarily always higher than that in the treated group. Although the sHR can affect the comparison of CIFs, i.e., sHR > 1 can indicate that $CIF_1 > CIF_0$ and vice versa, it reflects the relative change in the instantaneous rates of occurrence of a given type of event in subjects who have not yet experienced that event between two groups. Researchers may find it difficult to interpret the results when individuals who had a competing event are retained in the risk set (8). 2) Both the cause-specific Cox model and Fine-Gray model depend on an assumption of the proportionality of the CSH and the SDH; as a consequence, many published survival analyses report only a single cHR or sHR, which is an average of specific hazard ratios at different time points. However, if the above assumption is violated, a single HR is difficult to interpret because the true HR varies over time. 3) Because of the semi-parametric nature of the two regression models, the "relative" hazard rates cHR and sHR are not interchangeable with the "absolute" hazard rate without baseline hazards, which may make their clinical interpretation difficult to conceptualize.

Considering the above limitation, especially the problem of clinical interpretation, some researchers recommended an alternative statistic (9-11): restricted mean time lost (RMTL). RMTL can be estimated as the area under the CIF curve up to a specified time point and interpreted as the mean time lost due to a specific cause during a predefined time window. Thus, compared to that of HRs, the clinical interpretation of the RMTL, which is based on a time scale, can easily be understood by doctors and patients (12-14). The difference in RMTL (RMTLd) is used to qualify the treatment effect and is also directly associated with comparisons of CIFs.

Although Anderson (9) and Zhao (10) introduced the concept of RMTL, neither of them discussed the corresponding estimation and hypothetical test based on the RMTLd. Lyu (11) presented a statistical inference framework and sample size estimator based on the RMTLd, but it seemed to be relatively conservative based on simulations. Therefore, in this article, we introduce a new RMTLd-based statistical inference framework and sample size formula and demonstrate its performance through simulation and illustrative examples.

**METHODS**

Without loss of generality, only one event of interest ( $j = 1$ ) and one competing event ( $j = 2$ ) are assumed.   $T$   is defined as the observed time (time to event or censoring time).

**Estimation of the RMTLd**

The nonparametric estimation of the CIF is as follows:

$$\hat{F}_j(t) = \sum_{t_i} (d_{ij} / Y(t_i)) \hat{S}(t_{i-1})$$

where   $t_i$   is the $i$th   ordered event time,   $d_{ij}$   is the number of events of cause $j$ that occur at time   $t_i$,   $Y(t_i)$   is the number of subjects at risk at time   $t_i$, and   $S(t)$   is the event-free survival probability.   $\tau$   is the chosen time point, and   $\tau \leq T$ . For simplicity, we denote the RMTL of the event of interest to be   $\mu = \int_0^\tau F_1(t)dt$ ; then, the nonparametric estimation of   $\mu$   is given by

$$\hat{\mu} = \int_0^\tau \hat{F}_1(t)dt = \sum_{t_i \leq \tau} (d_{i1} / Y(t_i))\hat{S}(t_{i-1})(t_i - t_{i-1})$$

which can be interpreted as the mean time lost due to a specific cause within the $\tau$ year window. The variance in $\hat{\mu}$ can be estimated based on the derivation of the martingale approximation (15) (for the detailed process, see Web Appendix 1):

$$\text{var}(\hat{\mu}) = \int_0^\tau \left\{ (\tau - t)\frac{1 - \hat{F}_2(t)}{Y(t)} - \frac{1}{Y(t)}\int_t^\tau \hat{F}_1(u)du \right\}^2 \frac{Y(t)}{\hat{S}(t)}d\hat{F}_1(t)$$

$$+ \int_0^\tau \left\{ (\tau - t)\frac{\hat{F}_1(t)}{Y(t)} - \frac{1}{Y(t)}\int_t^\tau \hat{F}_1(u)du \right\}^2 \frac{Y(t)}{\hat{S}(t)}d\hat{F}_2(t)$$

Let $\hat{\mu}_k(\tau)$ be the RMTL of the event of interest in group $k\left(k = 0, 1\right)$; then, $\hat{\mu}_k(\tau)$ denotes the estimated RMTL, and $\text{var}(\hat{\mu}_k(\tau))$ corresponds to the variance in $\hat{\mu}_k(\tau)$. Then, the RMTLd between two groups is $\hat{\Delta} = \hat{\mu}_1(\tau) - \hat{\mu}_0(\tau)$, and the corresponding variance is $\text{var}(\hat{\Delta}) = \text{var}(\hat{\mu}_1(\tau)) + \text{var}(\hat{\mu}_0(\tau))$. In large samples, the $100(1 - \alpha)\%$ confidence interval (CI) of the RMTLd is estimated as

$$\hat{\Delta} \pm z_{\alpha/2}\sqrt{\text{var}(\hat{\Delta})} ,$$

where $z_\alpha$ is the upper $100\alpha\%$ quantile of the standard normal distribution.

**Hypothetical test**

The null and alternative hypotheses of the RMTLd test are

$$H_0 : \Delta = \mu_1(\tau) - \mu_0(\tau) = 0 \quad \text{and} \quad H_1 : \Delta = \mu_1(\tau) - \mu_0(\tau) \neq 0 , \text{ respectively.}$$

Under the null hypothesis $H_0$, the RMTLd test statistic can be computed as

$$Z_d = \frac{\hat{\Delta}}{\sqrt{\text{var}(\hat{\Delta})}} ,$$

which asymptotically follows a standard normal distribution.

**Sample size**

Suppose $n_0$ and $n_1$ are the required sample sizes in the control group and treatment group, respectively, and $r = n_1/n_0$ is the ratio of sample sizes. Then, assume we test the null hypothesis with power $1-\beta$ at a two-sided significance level $\alpha$. Under alternative hypothesis $H_1$, we then have

$$1-\beta = \Phi\left\{\left[\frac{\Delta}{\sqrt{\sigma_0^2/n_0 + \sigma_1^2/n_1}}\right] - z_{1-\alpha/2}\right\}$$

Hence, the total sample size is (for the detailed derivation, see Web Appendix 2)

$$n = \frac{(1+r)(z_{1-\beta} + z_{1-2/\alpha})^2}{\Delta^2/(\sigma_0^2 + \sigma_1^2 r^{-1})},$$

$z_p = \Phi^{-1}(p)$ is the inverse standard normal distribution function at probability $p$, and the population variance $\sigma_k^2$ of group $k$ can be estimated as $\hat{\sigma}_k^2 = n_k^* \operatorname{var}^*(\hat{\mu}_k(\tau))$, where $n_k^*$ and $\operatorname{var}^*(\hat{\mu}_k(\tau))$ can be obtained through a pilot or previous study.

**Simulation setup**

In the simulation setup, we assessed the performance of the estimation of the RMTLd, the RMTLd test and the RMTLd-based sample size under different scenarios: 1) no difference between groups (Figure 1A); 2) a proportional SDH with sHR $\approx 0.905$ (Figure 1B); 3) a proportional SDH with sHR $\approx 0.741$ (Figure 1C); 4) an early difference between groups (Figure 1D); 5) a late difference with curves separated at $t = 1$ year (Figure 1E); and 6) a late difference with curves separated at $t = 2$ years (Figure 1F).

Let the type of interest and competing events be generated through the binomial distributions $B(N, p_1)$ and $B(N, 1-p_1)$, where $N$ is defined as the sample size of each group and $p_1$ is the maximum cumulative incidence of events of interest, which is set to $p_1 = 0.7$. The parameter settings of failure time $T_j$ ($j = 1, 2$) correspond to the event of interest and the competing event, respectively) under different situations are shown in Web Table 1, and the censoring times of the two groups are based on the uniform distributions $U(0, a)$ and $U(0, b)$, respectively. Next, define the observed time $T = \min(T_j, C)$ and the event indicator $\delta_j = I(T_j > C)$. The censoring rates are required to be similar between the two groups and are approximately 0%, 15%, 30% or 45% by changing the settings of $a$ and $b$. For the sample size, we consider both a balanced design ($n_0 = n_1 = 300, 500, 1000$) and an unbalanced design ($n_0 = 300, n_1 = 500; n_0 = 500, n_1 = 1000$). For all scenarios, a nominal level $\alpha = 0.05$ is applied, and the specific time point $\tau$ is selected as the minimum of the maximum follow-up time of the two groups (16). All simulations are performed using 10,000 replications.

To evaluate the performance of the RMTLd estimation, we determined the true RMTLd at $\tau = 4$ years with a total sample size of n = 1,000,000 ($n_0 = n_1$) under the different scenarios. The true RMTLd between groups for the event of interest under scenarios A-F in Figure 1 are 0.00004, -0.3935, -0.5141, -0.2986, -0.3517 and -0.1729 years, respectively, over a period of 4 years. Then, according to the above settings, we sampled from this large sample to calculate the mean relative bias (Rel bias), the mean square error (RMSE), relative standard error (Rel SE) and coverage (Cov) of the RMTLd (17) to measure the performance of the estimation of the RMTLd.

Meanwhile, we compared the Type I error and statistical power of the Gray test and proposed the RMTLd test to evaluate the performance of the RMTLd test. To evaluate the type I error rate, the CIFs of the events of interest and competing events were assumed to be $F_1(t) = p_1\{1 - \exp(-t)\}$ and $F_2(t) = (1 - p_1)\{1 - \exp(-t)\}$, respectively, so the failure time in both groups was generated from $\Pr(T_j \leq t \mid J = j) = F_j(t) / \Pr(J = j)$, given the event type $J = j(j = 1, 2)$, as shown in Figure 1A.

To assess the statistical power, several situations were considered (Figure 1B -F): 1) The proportional SDH assumption is met: failure times were generated from the CIFs (18) $F_1(t \mid Z) = 1 - [1 - p_1(1 - e^{-t})]^{\exp(\theta Z)}$ and $F_2(t \mid Z) = (1 - p_1)^{\exp(\theta Z)}(1 - e^{-t \exp(\theta Z)})$, where $Z$ is the group indicator ($Z = 0$ and 1 for the control group and treatment group, respectively). Meanwhile, we considered two scenarios, sHR $\approx 0.905$ and sHR $\approx 0.741$, corresponding to Figure 1B and Figure 1C, respectively; 2) The proportional SDH assumption is violated: both the early difference (Figure 1D) and the late difference (Figure 1E - F) in the CIFs were considered. The failure time was generated based on CIFs with piecewise Weibull distributions $W(\lambda, \kappa)$ (where $\lambda$ and $\kappa$ are the scale parameter and shape parameter, respectively): $F_1(t) = p_1(1 - \exp(-(t / \lambda)^\kappa))$ and $F_2(t) = (1 - p_1)(1 - \exp(-(t / \lambda)^\kappa))$ (19). The specific parameter settings of all scenarios are presented in Web Table 1.

To evaluate the performance of the proposed sample size estimation, we set $\alpha = 0.05$ and $\beta = 0.20$ (the targeted power was 80%) and generated the necessary parameters by averaging over each simulation to calculate the RMTLd-based sample sizes under different situations (Figure 1B - F). Next, we simulated the observed power of the Gray test and the RMTLd test based on the calculated sample sizes through 10,000 simulations.

## RESULTS

### Estimation of the RMTLd

The results for the performance criterion of the estimation of the RMTLd are summarized in Table 1 and Table 2. Considering that the true RMTLd in scenario A is approximately equal to 0, we replaced the mean relative bias (Rel bias) with bias to assess the performance (20). In summary, the estimation of RMTLd has a small bias (or the mean relative bias) under all scenarios, and the root mean square error decreases with increasing sample size and decreasing censoring. Meanwhile, the relative standard error is approximately equal to 1, and the coverage falls within a reasonable range.

**Hypothetical test**

For each scenario (Figure 1A - F), the type I error rate and statistical power results are summarized in Table 3. The type I error rates in Table 3 show that both the proposed RMTLd test and the Gray test have well-controlled error rates. Under the proportional SDH assumption (B and C), the RMTLd test has similar power to the Gray test. In the early difference scenario (D), the RMTLd test provides significantly greater power than the Gray test. Meanwhile, as the censoring rate increases, the power of the two tests increases. This tendency may be interpreted as follows: because of high censoring, the number of patients in the later part of the CIF curve may be small, resulting in increased variability in the shape of the curve. The results for the late difference situations (E and F) show that the power of the two tests increases as the sample size increases and when the difference is larger and that the RMTLd test provides higher (or much higher) values than the Gray test. According to the above findings, the RMTLd test has relatively robust performance in different situations.

**Sample size**

For each scenario (Figure B - F), 10,000 simulations were performed to evaluate the observed power of the Gray test and the RMTLd test under the RMTLd-based sample size, and the results are shown in Table 4. Under proportional subdistribution hazards scenarios (B and C), the power of the RMTLd test and the Gray test is approximately equal to the predefined level of 80%. In the early difference scenario (D), the power of the RMTLd test is larger than the prespecified level, while that of the Gray test is much lower than 80%. In the late difference scenarios (E and F), the observed power of the RMTLd test is close to 80%, but that of the Gary test has an obvious decrease with a smaller difference (F).

In summary, the sample size based on the RMTLd can obtain a nominal power of approximately 80%, except in the early difference scenario. Therefore, the validity of the RMTLd-based sample size should be acceptable.

**Illustrative examples**

Example 1. Data from 599 American Chinese patients with cervical cancer diagnosed between 1988 and 2008 were obtained from the Surveillance, Epidemiology, and End Results (SEER) database of the National Cancer Institute to assess the association of surgical factors on survival. In our analysis, death from cervical cancer was defined as the event of interest, while death from other causes was defined as a competing event (21, 22). The rate of the event of interest in the non-surgery group (n = 101) was 29.70%, while the rate was 5.82% in the surgery group (n = 498). The corresponding censoring rates were 38.61% and 81.73%, respectively. Figure 2A shows the CIF curve of the event of interest, and Table 5 shows the statistical results of different tests.

The CSH-based results suggested a positive association of surgery on death from cervical cancer (cHR = 0.132, 95% CI: 0.079 to 0.220), and the assumption of the proportionality of the CSH was satisfied ($P = 0.596$). Meanwhile, the SDH-based method showed a positive association of surgery on the event of interest (sHR = 0.158, 95% CI: 0.095 to 0.262). A test of the proportional SDH assumption yielded a result of $P = 0.230$. Due to the semi-parametric nature of the regression model, neither the CSH nor the SDH could be obtained in any group, resulting in empty cells in Table 5.

Next, we let $\tau_1$ = 25.667 years, which corresponds to the shortest maximum follow-up time between the two groups. Table 5 shows that the RMTL of the non-surgery group was 7.485 years, while that of the surgery group was 1.346 years. The results can be interpreted as follows: during 25.667 ($\tau_1$) years of follow-up, the mean loss of life due to death from cervical cancer of the patients in the non-surgery and surgery groups was 7.485 and 1.346 years, respectively. The RMTLd test results also favor the surgery group (RMTLd = -6.139, 95% CI: -8.400 to -3.878), and the RMTLd = -6.139 years (the RMTL of the surgery group minus the RMTL of the non-surgery group) indicated that the patients without surgery lost an additional 6.139 years of life due to cervical cancer within 25.667 ($\tau_1$) years of follow-up. Thus, the RMTLd-based results provided a more acceptable conception of the time scale.

Example 2. A total of 2,279 patients with acute lymphoid leukemia who received allogeneic bone transplants from a human leukocyte antigen (HLA)-identical sibling donor were recorded in the European Group for Blood and Marrow Transplantation (23). We studied the association of donor-recipient gender match on survival, so a total of 2,279 patients were grouped into gender mismatched (n = 545) and matched (n = 1734) groups. Death after transplantation was documented as the event of interest, and relapse from transplantation was a competing event. The proportions of the event of interest in the mismatched and matched groups were 26.61% and 22.38%, and the censoring rates were 56.88% and 61.48%, respectively.

The results based on the CSH showed no significant difference between the two groups (cHR = 0.828, 95% CI: 0.684 to 1.002), and the proportional CSH assumption was violated ($P$ = 0.001). The result based on the SDH also indicated no statistically significant difference (sHR = 0.835, 95% CI: 0.692 to 1.008), and the proportional SDH assumption was not satisfied ($P$ = 0.004). Regarding the above results, the true cHR and sHR may vary with time rather than be constant (cHR = 0.828 and sHR = 0.835), which makes clinical interpretation difficult.

Unlike the above CSH-based and SDH-based tests, the RMTLd test detected a difference between the two groups and showed a positive association in the matched group over $\tau_2$ = 16.238 years of follow-up (RMTLd = -1.023, 95% CI: -1.755 to -0.291), and the RMTLd indicated that the gender-mismatched patients lost an additional 1.023 years of life on average during the 16.238 years. This significant result of the RMTL test was not unexpected because this example (Figure 2B) corresponds to simulation scenario F (Figure 1F), and it showed that the RMTLd test had higher power than the Gray test, as shown in Table 3.

Example 3. The Adaptive COVID-19 Treatment Trial (ACTT-1) is a placebo-controlled trial to assess remdesivir use in patients hospitalized with COVID-19 (2, 24). The data were reconstructed (for the detailed process, see Web Appendix 3) because the original data were not publicly available (2); the event of interest was defined as recovery, and the corresponding competing event was death. In ACTT-1, 541 patients were assigned to the remdesivir group, and 521 were assigned to the placebo group. The proportions of recovered patients in the remdesivir and control groups were 70.98% and 63.92%, respectively, and the censoring rates were 17.56% and 20.92%, respectively. Figure 3A shows the CIF curve of recovery between groups.

The results based on the CSH and SDH (Table 5) showed significant differences, and the proportional CSH assumption was satisfied ($P$ = 0.056), while the SDH assumption was violated ($P$ = 0.002). In regard to the RMTL, we note that different from the event of interest in example 1 (in which death from cervical cancer was a negative outcome), the event of interest in this example, i.e., recovery, was a positive outcome. Thus, a larger RMTL indicated better therapy. From Table 5, the RMTLs of the placebo and remdesivir groups were 10.859 and 13.286 days, respectively, which can be interpreted to indicate that over the 28 ($\tau_3$) days of follow-up, the patients in the placebo group had 10.859 postrecovery days, on average, while the patients in the remdesivir group had 13.286 days. In other words, the patients in the placebo and remdesivir groups had been recovered for an average of 10.859 and 13.286 days by day 28, respectively. The RMTLd = 2.427 days (95% CI: 1.242 to 3.612) favored the remdesivir group and showed that the patients in the remdesivir group recovered 2.427 days earlier than those in the placebo group in the 28-day period.

Meanwhile, due to the clinical significance of COVID-19 research, we set $\tau_3$ = 28 days and re-estimated that the sample sizes based on the sHR (25) and the RMTLd were 658 and 517, respectively. Moreover, based on different time points, we calculated the RMTLd-based sample sizes. As Figure 3B shows, the RMTLd-based sample sizes were always smaller than the sHR-based sample sizes.

## DISCUSSION

The presence of competing risks makes treatment effect assessment in clinical trials and epidemiological studies with time-to-event endpoints more cumbersome. The commonly reported quantitative measures are the cHR and sHR, where the former might be used to study the etiology of diseases from biological mechanisms and the latter might be more suitable for predicting an individual's risk of a specific outcome (7).

However, based on our examples, there are still some limitations to the above two indicators based on HR. First, as a "relative" measure, HRs (both the cHR and sHR) cannot be easily understood when a baseline hazard is lacking (e.g., of a control group), even though the proportional CSH and SDH assumptions were satisfied in example 1. Moreover, the cHR = 0.132 and sHR = 0.158 in example 1 cannot be directly interpreted, as the "risk" of death from cervical cancer decreased by 86.8% or 84.2%, respectively, for the surgery group; rather, this result should be understood as an 86.8% or 84.2% decrease, respectively, in the "hazard" of death from cervical cancer, which is difficult to interpret clinically (2, 6). Furthermore, because the proportional assumptions were violated in example 2, the CSH and SDH curves of the two groups in Web Figure 1 (obtained through the nonparametric technique) have a late difference, showing that the cHRs and sHRs may vary over time. Therefore, a weighted average HR alone may fail to quantify and interpret the treatment effect.

As an alternative statistic, some researchers (9-11) developed the RMTL, which corresponds to the area under the CIF curve. Thus, the RMTL can easily be implemented and interpreted on a time scale. Meanwhile, as an "absolute" measure, the RMTLd can be used to supplement the cHR and sHR to evaluate the treatment effect. Moreover, the RMTLd-based test does not require any model assumptions.

Based on the RMTLd, we introduced a new statistical inference framework and sample size estimator. From our simulation results, the performance of the estimation of the RMTLd and the RMTLd test are acceptable and robust. However, notably, the simulation results of 45% censoring are not shown in Table 1 and Table 2 because we set the true RMTLd at $t = 4$ years; that is, the final follow-up time should be equal to or greater than 4 years for the generation of survival data, which is violated with 45% censoring (for more discussion, see the Web Table 2, Web Table 3 and Web Figure 2). In summary, the proposed RMTLd is accurate, and the RMTLd test has well-controlled type I error rates and has similar power to (or even larger power than) the Gray test. Meanwhile, the results of the sample size simulation showed that the power of the RMTLd-based sample size can approximately achieve the predefined level of power, and we also calculated the RMTLd-based and sHR-based sample sizes in examples 1 and 2 (regardless of the clinical significance, shown in Web Figure 3), which also indicates that the proposed sample size formula is effective and suitable. Thus, for clinical trials and epidemiological studies, the RMTLd test may be the most robust approach for competing risks.

However, there are still some limitations in this study: 1) the time point $\tau$ was simply restricted to be the shortest maximum follow-up time of the two groups, whereas from a practical perspective, $\tau$ can be chosen according to scientific clinical or epidemiological knowledge, e.g., 28 days based on COVID-19 studies; 2) the results of the statistical tests in the examples were applied to illustrate the analysis and interpret the outcomes, but they did not offer any clinically relevant conclusions.

In summary, in competing risks analysis, we recommend the RMTLd as a supplement to the cHR or sHR in the measurement of treatment effects when the proportional hazard assumptions are satisfied. When the assumptions are violated, the RMTLd could be selected as an alternative statistic for summarizing and interpreting the treatment effect.

## ACKNOWLEDGMENTS

This work was supported by the National Natural Science Foundation of China (82173622, 81903411, 81673268), the Guangdong Basic and Applied Basic Research Foundation (2019A1515011506) and Natural Science Foundation of Guangdong Province (2018A030313849).

Conflict of interest: none declared.

## REFERENCES

1. Dillman A, Park JJH, Zoratti MJ, et al. Reporting and design of randomized controlled trials for COVID-19: A systematic review. Contemp Clin Trials. 2021;101:106239.

2. McCaw ZR, Tian L, Vassy JL, et al. How to Quantify and Interpret Treatment Effects in Comparative Clinical Studies of COVID-19. Ann Intern Med. 2020;173(8):632-637.

3. Andersen PK, Geskus RB, de Witte T, et al. Competing risks in epidemiology: possibilities and pitfalls. Int J Epidemiol. 2012;41(3):861-70.

4. Schuster NA, Hoogendijk EO, Kok AAL, et al. Ignoring competing events in the analysis of survival data may lead to biased results: a nonmathematical illustration of competing risk analysis. J Clin Epidemiol. 2020;122:42-48.

5. Latouche A, Allignol A, Beyersmann J, et al. A competing risks analysis should report results on all cause-specific hazards and cumulative incidence functions. J Clin Epidemiol. 2013;66(6):648-53.

6. Sutradhar R, Austin PC. Relative rates not relative risks: addressing a widespread misinterpretation of hazard ratios. Ann Epidemiol. 2018;28(1):54-57.

7. Lau B, Cole SR, Gange SJ. Competing risk regression models for epidemiologic data. Am J Epidemiol. 2009;170(2):244-56.

8. Austin PC, Fine JP. Practical recommendations for reporting Fine-Gray model analyses for competing risk data. Stat Med. 2017;36(27):4391-4400.

9. Andersen PK. Decomposition of number of life years lost according to causes of death. Stat Med. 2013;32(30):5278-85.

10. Zhao L, Tian L, Claggett B, et al. Estimating Treatment Effect With Clinical Interpretation From a Comparative Clinical Trial With an End Point Subject to Competing Risks. JAMA Cardiol. 2018;3(4):357-358.

11. Lyu J, Hou Y, Chen Z. The use of restricted mean time lost under competing risks data. BMC Med Res Methodol. 2020;20(1):197.

12. Royston P, Parmar MK. Restricted mean survival time: an alternative to the hazard ratio for the design and analysis of randomized trials with a time-to-event outcome. BMC Med Res Methodol. 2013;13:152.

13. Pak K, Uno H, Kim DH, et al. Interpretability of Cancer Clinical Trial Results Using Restricted Mean Survival Time as an Alternative to the Hazard Ratio. JAMA Oncol. 2017;3(12):1692-1696.

14. Staerk L, Preis SR, Lin H, et al. Novel Risk Modeling Approach of Atrial Fibrillation

With Restricted Mean Survival Times: Application in the Framingham Heart Study Community-Based Cohort. Circ Cardiovasc Qual Outcomes. 2020;13(4):e005918.

15. Bajorunaite R, Klein J. Two-sample tests of the equality of two cumulative incidence function. Comput Stat Data Anal. 2007;51(9):4269-4281.

16. Tian L, Jin H, Uno H, et al. On the empirical choice of the time window for restricted mean survival time. Biometrics. 2020;76(4):1157-1166.

17. Morris TP, White IR, Crowther MJ. Using simulation studies to evaluate statistical methods. Stat Med. 2019;38(11):2074-2102.

18. Beyersmann J, Allignol A, Schumacher M. Competing Risks and Multistate Models with R. Springer, New York, NY, 2012.

19. Li J, Le-Rademacher J, Zhang MJ. Weighted comparison of two cumulative incidence functions with R-CIFsmry package. Comput Methods Programs Biomed. 2014;116(3):205-14.

20. Burton A, Altman DG, Royston P, Holder RL. The design of simulation studies in medical statistics. Stat Med. 2006;25(24):4279-92.

21. Wasif N, Neville M, Gray R, et.al. Competing Risk of Death in Elderly Patients with Newly Diagnosed Stage I Breast Cancer. J Am Coll Surg. 2019;229(1):30-36.e1.

22. Lee M, Feuer EJ, Fine JP. On the analysis of discrete time competing risks data. Biometrics. 2018;74(4):1468-1481.

23. de Wreede LC, Fiocco M, Putter H. mstate: An R Package for the Analysis of Competing Risks and Multi-State Models. J Stat Softw. 2011;38(7):1-30.

24. Beigel JH, Tomashek KM, Dodd LE, et.al. Remdesivir for the Treatment of Covid-19 - Final Report. N Engl J Med. 2020;383(19):1813-1826.

25. Tai BC, Wee J, Machin D. Analysis and design of randomised clinical trials involving competing risks endpoints. Trials. 2011;12:127.

26. Lyu J, Chen J, Hou Y, Chen Z. Comparison of two treatments in the presence of competing risks. Pharm Stat. 2020;19(6):746-762.

## Table 1 Results of the estimation of the RMTLd under proportional SDH

| $(n_0, n_1)$ | CR (%) | A | | | | B | | | | C | | | |
|---|---|---|---|---|---|---|---|---|---|---|---|---|---|
| | | Bias[a] ($\times10^{-2}$) | RMSE | Rel SE | Cov[b] | Rel bias[a] ($\times10^{-2}$) | RMSE | Rel SE | Cov[b] | Rel bias[a] ($\times10^{-2}$) | RMSE | Rel SE | Cov[b] |
| (300, 300) | 0 | -0.0437 | 0.1289 | 1.0012 | 0.9503 | 0.2837 | 0.1324 | 0.9887 | 0.9441 | -0.1210 | 0.1291 | 1.0023 | 0.9509 |
| | 15 | 0.1275 | 0.1372 | 0.9945 | 0.9487 | -0.2349 | 0.1378 | 1.0004 | 0.9526 | 0.0310 | 0.1365 | 0.9959 | 0.9469 |
| | 30 | -0.1657 | 0.1543 | 0.9819 | 0.9456 | -1.4239 | 0.1543 | 0.9852 | 0.9471 | -2.6107 | 0.1502 | 0.9927 | 0.9476 |
| (500, 500) | 0 | -0.1462 | 0.1003 | 0.9977 | 0.9477 | -0.2314 | 0.1018 | 0.9969 | 0.9469 | -0.1333 | 0.1020 | 0.9837 | 0.9432 |
| | 15 | 0.1584 | 0.1049 | 1.0080 | 0.9537 | -0.2476 | 0.1064 | 1.0044 | 0.9519 | -0.4303 | 0.1059 | 0.9954 | 0.9482 |
| | 30 | -0.0464 | 0.1180 | 0.9985 | 0.9503 | -1.9034 | 0.1196 | 0.9894 | 0.9508 | -2.0862 | 0.1186 | 0.9763 | 0.9462 |
| (1000,1000) | 0 | 0.0463 | 0.0718 | 0.9865 | 0.9484 | 0.1519 | 0.0716 | 1.0029 | 0.9493 | 0.0659 | 0.0708 | 1.0030 | 0.9540 |
| | 15 | -0.0825 | 0.0744 | 1.0055 | 0.9499 | -0.6177 | 0.0746 | 1.0140 | 0.9516 | -0.4003 | 0.0738 | 1.0108 | 0.9491 |
| | 30 | 0.1286 | 0.0850 | 0.9846 | 0.9449 | -1.3734 | 0.0855 | 0.9820 | 0.9459 | -1.7447 | 0.0827 | 0.9949 | 0.9470 |
| (300, 500) | 0 | 0.2184 | 0.1162 | 0.9943 | 0.9499 | -0.4228 | 0.1164 | 1.0036 | 0.9501 | -0.4984 | 0.1159 | 0.9989 | 0.9475 |
| | 15 | -0.0496 | 0.1229 | 0.9923 | 0.9472 | -0.2165 | 0.1236 | 0.9952 | 0.9495 | -0.1835 | 0.1220 | 0.9982 | 0.9490 |
| | 30 | 0.4669 | 0.1366 | 0.9935 | 0.9477 | -1.7534 | 0.1380 | 0.9861 | 0.9455 | -2.4770 | 0.1376 | 0.9750 | 0.9439 |
| (500, 1000) | 0 | -0.0903 | 0.0861 | 1.0072 | 0.9505 | 0.0500 | 0.0866 | 1.0107 | 0.9503 | 0.2498 | 0.0872 | 0.9952 | 0.9451 |
| | 15 | -0.0614 | 0.0904 | 1.0133 | 0.9527 | -0.4181 | 0.0931 | 0.9914 | 0.9493 | -0.4221 | 0.0911 | 1.0038 | 0.9506 |
| | 30 | 0.1896 | 0.1043 | 0.9799 | 0.9467 | -1.7623 | 0.1029 | 0.9960 | 0.9480 | -2.1686 | 0.1023 | 0.9897 | 0.9444 |

Abbreviations: Rel bias, the mean relative bias, the mean bias relative to true RMTLd; RMSE, root of mean-squared error; Rel SE, relative standard error, average model standard error/the empirical standard error; Cov, coverage, CR, censoring rate.

[a]: The true RMTLd under scenario A-C are 0.00004, -0.3935 and -0.5141 years during 4 years, respectively.

[b]: The reasonable coverage (0.9457, 0.9543) is based on $(0.95 \pm 1.96\sqrt{0.95(1-0.95)/10000})$ (20).

Table 2 Results of the estimation of the RMTLd under non-proportional SDH

| $(n_0, n_1)$ | CR (%) | D | | | | E | | | | F | | | |
|---|---|---|---|---|---|---|---|---|---|---|---|---|---|
| | | Rel bias[a] ($\times 10^{-2}$) | RMSE | Rel SE | Cov[b] | Rel bias[a] ($\times 10^{-2}$) | RMSE | Rel SE | Cov[b] | Rel bias[a] ($\times 10^{-2}$) | RMSE | Rel SE | Cov[b] |
| (300, 300) | 0 | 0.2004 | 0.1032 | 1.0206 | 0.9528 | -0.4009 | 0.1062 | 1.0071 | 0.9523 | -0.0479 | 0.1085 | 0.9898 | 0.9444 |
| | 15 | 0.6546 | 0.1120 | 0.9962 | 0.9477 | -0.2764 | 0.1119 | 0.9967 | 0.9487 | 0.7193 | 0.1112 | 1.0107 | 0.9512 |
| | 30 | 0.2601 | 0.1182 | 1.0034 | 0.9512 | -0.4848 | 0.1170 | 1.0055 | 0.9519 | -0.4682 | 0.1182 | 1.0061 | 0.9486 |
| (500, 500) | 0 | -0.2143 | 0.0818 | 0.9985 | 0.9493 | 0.0281 | 0.0835 | 0.9931 | 0.9485 | 0.1572 | 0.0836 | 0.9954 | 0.9488 |
| | 15 | 0.6183 | 0.0861 | 1.0050 | 0.9517 | -0.2761 | 0.0857 | 1.0089 | 0.9516 | 0.2146 | 0.0865 | 1.0068 | 0.9513 |
| | 30 | 0.7100 | 0.0928 | 0.9910 | 0.9471 | -0.0503 | 0.0912 | 0.9992 | 0.9499 | 0.6477 | 0.0930 | 0.9912 | 0.9479 |
| (1000,1000) | 0 | -0.0850 | 0.0576 | 1.0033 | 0.9511 | -0.0996 | 0.0591 | 0.9915 | 0.9453 | -0.0263 | 0.0588 | 1.0011 | 0.9502 |
| | 15 | 0.0271 | 0.0610 | 1.0038 | 0.9527 | -0.2364 | 0.0612 | 0.9999 | 0.9526 | 0.1664 | 0.0617 | 0.9985 | 0.9483 |
| | 30 | 0.4661 | 0.0655 | 0.9927 | 0.9486 | 0.2138 | 0.0645 | 1.0000 | 0.9483 | 0.0847 | 0.0656 | 0.9946 | 0.9493 |
| (300, 500) | 0 | -0.3262 | 0.0968 | 1.0032 | 0.9490 | -0.0204 | 0.0953 | 0.9964 | 0.9480 | 0.3601 | 0.0959 | 0.9937 | 0.9492 |
| | 15 | -0.0674 | 0.1018 | 1.0084 | 0.9505 | -0.5680 | 0.0988 | 1.0065 | 0.9521 | 0.0784 | 0.1001 | 0.9986 | 0.9502 |
| | 30 | -0.2966 | 0.1096 | 0.9932 | 0.9467 | 0.9197 | 0.1063 | 0.9905 | 0.9462 | 1.3494 | 0.1063 | 0.9980 | 0.9498 |
| (500, 1000) | 0 | 0.0673 | 0.0737 | 0.9992 | 0.9510 | 0.1859 | 0.0705 | 1.0091 | 0.9498 | -0.3422 | 0.0712 | 1.0010 | 0.9494 |
| | 15 | 0.3837 | 0.0791 | 0.9824 | 0.9463 | -0.5905 | 0.0751 | 0.9924 | 0.9489 | 0.5657 | 0.0749 | 0.9992 | 0.9490 |
| | 30 | 0.4481 | 0.0822 | 1.0033 | 0.9487 | 0.2771 | 0.0787 | 1.0031 | 0.9507 | 0.4813 | 0.0793 | 1.0029 | 0.9518 |

Abbreviations: Rel bias, the mean relative bias, the mean bias relative to true RMTLd; RMSE, root of mean-squared error; Rel SE, relative standard error, average model standard error/the empirical standard error; Cov, coverage, CR, censoring rate.

[a]The true RMTLd under scenario D-F are -0.2986, -0.3517 and -0.1729 years during 4 years, respectively.

[b]The reasonable coverage (0.9457, 0.9543) is based on $(0.95 \pm 1.96\sqrt{0.95(1-0.95)/10000})$ (20).

Table 3 Type I error and statistical power of the Gary test and RMTLd test

| $(n_0, n_1)$ | CR (%) | Type I error[a] | | Statistical Power | | | | | | | | | |
| --- | --- | --- | --- | --- | --- | --- | --- | --- | --- | --- | --- | --- | --- |
| | | A | | B | | C | | D | | E | | F | |
| | | Gray | RMTLd | Gray | RMTLd | Gray | RMTLd | Gray | RMTLd | Gray | RMTLd | Gray | RMTLd |
| (300, 300) | 0 | 0.0497 | 0.0500 | 0.2343 | 0.2482 | 0.9468 | 0.9609 | 0.2236 | 0.6842 | 0.8706 | 0.9326 | 0.4714 | 0.4575 |
| | 15 | 0.0538 | 0.0531 | 0.2413 | 0.2366 | 0.9528 | 0.9486 | 0.3770 | 0.6812 | 0.8698 | 0.9039 | 0.4207 | 0.4056 |
| | 30 | 0.0501 | 0.0505 | 0.2226 | 0.2187 | 0.9412 | 0.9398 | 0.5604 | 0.6960 | 0.8487 | 0.8541 | 0.3487 | 0.3372 |
| | 45 | 0.0506 | 0.0504 | 0.1954 | 0.1969 | 0.9065 | 0.9074 | 0.7910 | 0.8464 | 0.7376 | 0.7281 | 0.1892 | 0.1952 |
| (500, 500) | 0 | 0.0504 | 0.0500 | 0.3497 | 0.3700 | 0.9945 | 0.9965 | 0.3374 | 0.8340 | 0.9441 | 0.9931 | 0.6222 | 0.6807 |
| | 15 | 0.0532 | 0.0525 | 0.3636 | 0.3472 | 0.9967 | 0.9961 | 0.5574 | 0.8344 | 0.9625 | 0.9870 | 0.5830 | 0.6224 |
| | 30 | 0.0477 | 0.0487 | 0.3453 | 0.3381 | 0.9959 | 0.9952 | 0.7739 | 0.8507 | 0.9677 | 0.9776 | 0.4963 | 0.5293 |
| | 45 | 0.0505 | 0.0510 | 0.3057 | 0.3039 | 0.9864 | 0.9867 | 0.9466 | 0.9674 | 0.9166 | 0.9218 | 0.2934 | 0.3249 |
| (1000,1000) | 0 | 0.0490 | 0.0505 | 0.6009 | 0.6087 | 1.0000 | 1.0000 | 0.5820 | 0.9694 | 0.9775 | 1.0000 | 0.8172 | 0.9411 |
| | 15 | 0.0474 | 0.0485 | 0.6185 | 0.5901 | 1.0000 | 1.0000 | 0.8459 | 0.9681 | 0.9922 | 0.9999 | 0.8221 | 0.9109 |
| | 30 | 0.0542 | 0.0528 | 0.6036 | 0.5866 | 1.0000 | 1.0000 | 0.9690 | 0.9731 | 0.9991 | 0.9999 | 0.7786 | 0.8488 |
| | 45 | 0.0477 | 0.0482 | 0.5353 | 0.5336 | 1.0000 | 1.0000 | 0.9989 | 0.9995 | 0.9972 | 0.9981 | 0.5136 | 0.5950 |
| (300, 500) | 0 | 0.0500 | 0.0513 | 0.2889 | 0.2957 | 0.9769 | 0.9837 | 0.2788 | 0.7277 | 0.8917 | 0.9697 | 0.5390 | 0.5509 |
| | 15 | 0.0515 | 0.0518 | 0.2954 | 0.2823 | 0.9839 | 0.9820 | 0.4642 | 0.7295 | 0.9096 | 0.9521 | 0.4853 | 0.4811 |
| | 30 | 0.0491 | 0.0523 | 0.2841 | 0.2715 | 0.9767 | 0.9753 | 0.6658 | 0.7444 | 0.9071 | 0.9171 | 0.3947 | 0.3963 |
| | 45 | 0.0501 | 0.0524 | 0.2461 | 0.2390 | 0.9567 | 0.9539 | 0.8825 | 0.8982 | 0.8133 | 0.8150 | 0.2167 | 0.2345 |
| (500, 1000) | 0 | 0.0511 | 0.0527 | 0.4577 | 0.4645 | 0.9996 | 0.9997 | 0.4374 | 0.8891 | 0.9373 | 0.9990 | 0.7010 | 0.8115 |
| | 15 | 0.0469 | 0.0478 | 0.4684 | 0.4406 | 0.9997 | 0.9996 | 0.6877 | 0.8822 | 0.9733 | 0.9982 | 0.6841 | 0.7533 |
| | 30 | 0.0526 | 0.0530 | 0.4517 | 0.4346 | 0.9995 | 0.9993 | 0.8794 | 0.9018 | 0.9904 | 0.9959 | 0.5912 | 0.6491 |
| | 45 | 0.0516 | 0.0507 | 0.3901 | 0.3795 | 0.9990 | 0.9985 | 0.9838 | 0.9845 | 0.9647 | 0.9732 | 0.3430 | 0.4128 |

Abbreviations: CR, Censoring Rate; Gray, Gray test; RMTLd, RMTLd test.

[a]The reasonable range (0.0457, 0.0543) based on the formula $(0.05 \pm 1.96\sqrt{0.05(1-0.05)/10000})$ (26).

Table 4 Observed power of the Gray and RMTLd tests based on the same sample[a]

| CR (%) | B | | | C | | | D | | | E | | | F | | |
|---|---|---|---|---|---|---|---|---|---|---|---|---|---|---|---|
| | N | Gray | RMTLd | N | Gray | RMTLd | N | Gray | RMTLd | N | Gray | RMTLd | N | Gray | RMTLd |
| 0 | 3180 | 0.7903 | 0.7885 | 370 | 0.7976 | 0.8380 | 1074 | 0.3567 | 0.8453 | 400 | 0.7470 | 0.7966 | 1308 | 0.7001 | 0.8009 |
| 15 | 3258 | 0.8282 | 0.7958 | 376 | 0.8238 | 0.8226 | 1060 | 0.5827 | 0.8472 | 438 | 0.7648 | 0.7808 | 1444 | 0.7174 | 0.7898 |
| 30 | 3348 | 0.8251 | 0.8102 | 372 | 0.7987 | 0.7943 | 980 | 0.7639 | 0.8477 | 496 | 0.7867 | 0.7778 | 1720 | 0.7114 | 0.7832 |
| 45 | 3600 | 0.7818 | 0.7798 | 428 | 0.7825 | 0.7833 | 566 | 0.7667 | 0.8300 | 650 | 0.7650 | 0.7618 | 3286 | 0.7134 | 0.8125 |

Abbreviations: CR, Censoring Rate; N, Total sample size; Gray, Gray test; RMTLd, RMTLd test.

[a]The sample sizes were calculated based on the RMTLd test. The chosen time point $\tau$ was the shortest maximum follow-up time of the two groups, and the pre-specified power was 80%.

Table 5.  Statistical results for the illustrative examples.

| Test | Example 1 ($\tau_1$ = 25.667 years) | | | | | Example 2 ($\tau_2$ = 16.238 years) | | | | | Example 3 ($\tau_3$ = 28 days) | | | | |
|---|---|---|---|---|---|---|---|---|---|---|---|---|---|---|---|
| | Non-surgery[ab] | Surgery[a] | Ratio/Difference[c] | 95% CI | $P$[d] | Mis-match[ab] | Match[a] | Ratio/Difference[c] | 95% CI | $P$[d] | Placebo[ab] | Rem-desivir[a] | Ratio/Difference[c] | 95% CI | $P$[d] |
| cHR | | | 0.132 | 0.079, 0.220 | <0.001 | | | 0.828 | 0.684, 1.002 | 0.051 | | | 1.326 | 1.145, 1.539 | <0.001 |
| sHR | | | 0.158 | 0.095, 0.262 | <0.001 | | | 0.835 | 0.692, 1.008 | 0.064 | | | 1.305 | 1.130, 1.506 | <0.001 |
| RMTLd | 7.485 | 1.346 | -6.139 | -8.400, -3.878 | <0.001 | 4.661 | 3.638 | -1.023 | -1.755, -0.291 | 0.006 | 10.859 | 13.286 | 2.427 | 1.242, 3.612 | <0.001 |

Abbreviations: cHR, Cause-specific Hazard Ratio; sHR, Subdistribution Hazard Ratio; RMTLd, the difference of Restricted Mean Time Lost.

[a] The cause-specific hazard, subdistribution hazard and RMTL in each group. Due to the semi-parametric nature, neither the CSH nor the SDH could be obtained in any group, resulting in empty cells in CSH and SDH.

[b] The reference group in cHR, sHR and RMTLd.

[c] The cHR and sHR are related to the corresponding ratio, while the RMTLd is related to the difference.

[d] *P*-values for the cHR, sHR and RMTLd were calculated by the log-rank test, Gray test and RMTLd test, respectively.

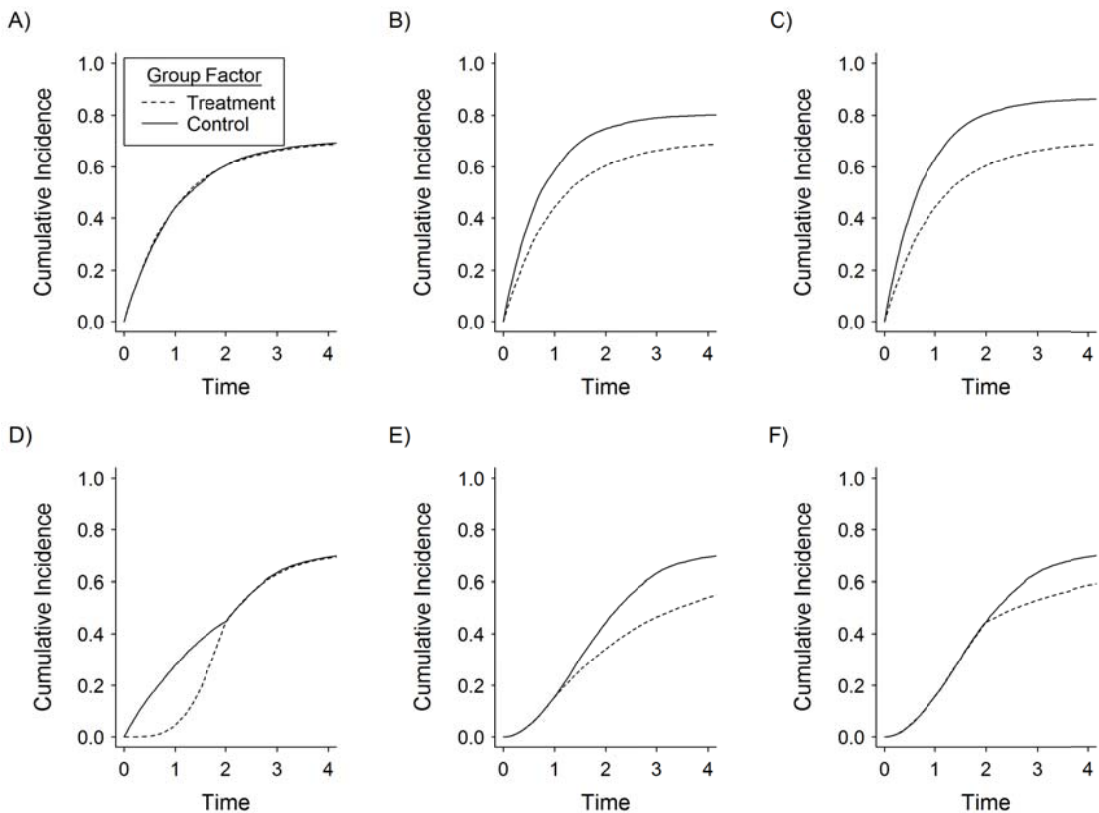

**Figure 1.** Scenarios considered in the simulation study. A) no difference in event of interest between groups; B) a proportional SDH with small difference; C) a proportional SDH with large difference; D) an early difference; E) a late difference with large difference; F) a late difference with small difference.

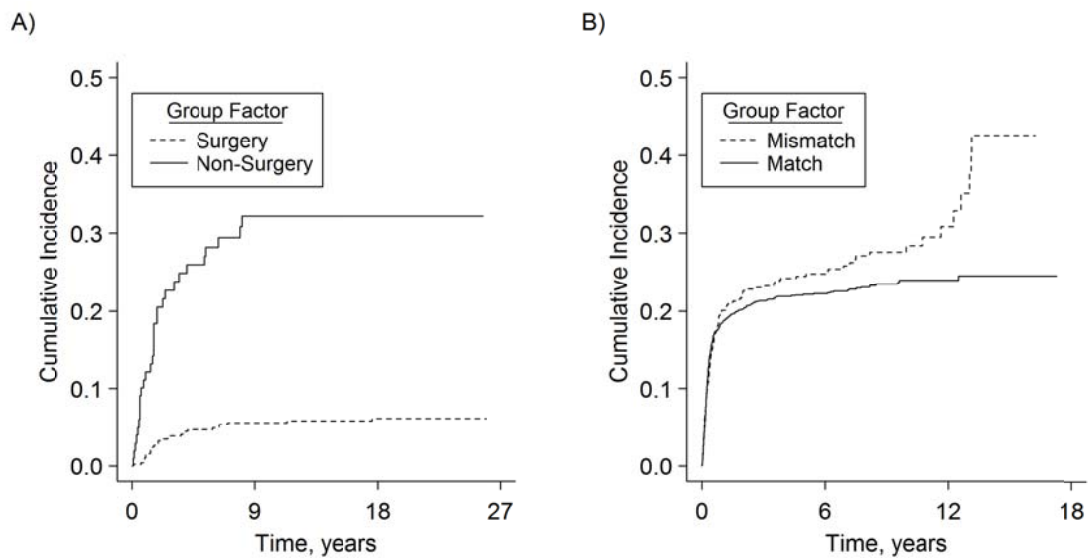

**Figure 2.** The cumulative incidence function of event of interest in example 1 (A) and example 2 (B). The restricted time point was chosen at 25.667 years in example 1 and 16.238 years in example 2.

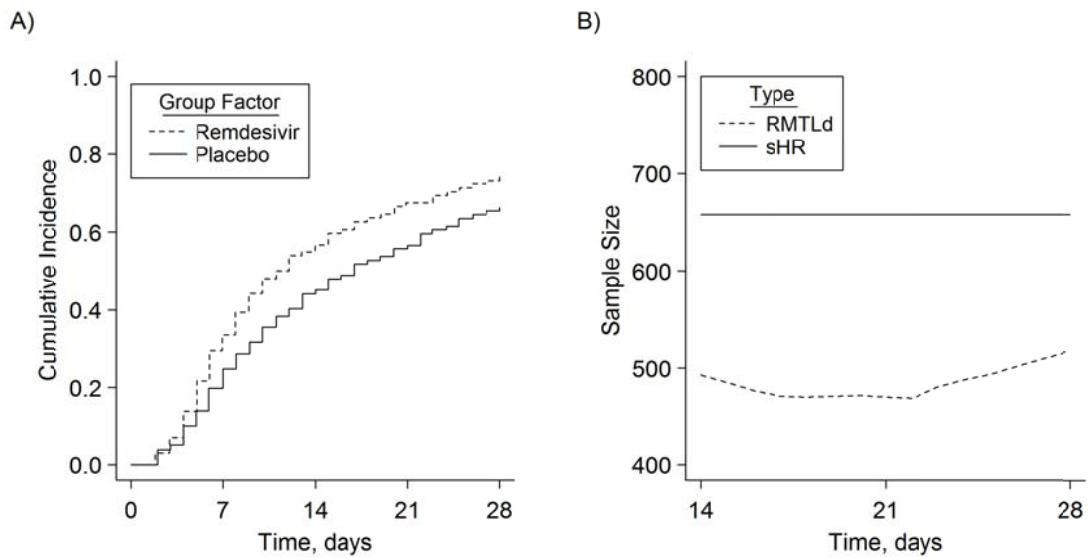

**Figure 3.** The cumulative incidence function of event of interest in example 3 (A) and sample size based on sHR and RMTLd (b). The restricted time points was chosen from 14 to 28 days when calculated RMTLd-based sample sizes.

**Web Material**

**Implementation of an alternative method for assessing competing risks:
restricted mean time lost**

Hongji Wu, Hao Yuan, Zijing Yang, Yawen Hou and Zheng Chen

**Web Appendix 1: The estimation of variance of RMTL**

Without loss of generality, we assume that interest event and competing event are recoded as $j = 1$ and $j = 2$ for the two groups ($k = 0, 1$), respectively. For the $i$th event timepoint, define $d_{ijk} = I(T = t_i, \delta = j, G = k)$ and $Y_k(t_i) = I(T \geq t_i, G = k)$ as the indicators of experiencing failure from cause $j$ by time $t_i$ and being at risk just before $t_i$ in group $k$, respectively. Then, the estimation of the cumulative incidence function of cause $j$ in group $k$ can be written as $\hat{F}_{jk}(t) = \sum_{t_i} (d_{ijk} / Y_k(t_i)) \hat{S}_k(t_{i-1})$, where $\hat{S}_k(t)$ is the Kaplan-Meier estimate when all events (both $j = 1$ and $j = 2$) in group $k$ are considered.

According to Bajorunaite (1), the integrated difference between the CIF of event of interest and the corresponding variance can be written as follows:

$$Z_p = \sqrt{n_0 n_1 / (n_0 + n_1)} \int_0^\tau \left\{ F_{11}(u) - F_{10}(u) \right\} du$$

and

$$\sigma_{Z_p}^2 = \frac{n_0 n_1}{n_0 + n_1} \sum_{k=0}^1 \left[ \int_0^\tau \left\{ (\tau - s) \frac{1 - F_{2k}(s)}{Y_k(s)} - \frac{1}{Y_k(s)} \int_s^\tau F_{1k}(u) du \right\}^2 \frac{Y_k(s)}{S_k(s)} dF_{1k}(s) \right.$$
$$\left. + \int_0^\tau \left\{ (\tau - s) \frac{F_{1k}(s)}{Y_k(s)} - \frac{1}{Y_k(s)} \int_s^\tau F_{1k}(u) du \right\}^2 \frac{Y_k(s)}{S_k(s)} dF_{2k}(s) \right]$$

where $n_k$ is the sample size in group $k$ and $\tau$ is the chosen time point. $Y_k(t)$ is the risk set at time $t$ in group $k$. Assuming that $\mu_k$ is the RMTL of event of interest in group $k$, formula 1 and formula 2 can be reduced to the following:

$$\hat{Z}_p = \sqrt{n_0 n_1 / (n_0 + n_1)} \int_0^\tau \left\{ \hat{F}_{11}(u) - \hat{F}_{10}(u) \right\} du$$
$$= \sqrt{n_0 n_1 / (n_0 + n_1)} \left( \int_0^\tau \hat{F}_{11}(u) du - \int_0^\tau \hat{F}_{10}(u) du \right),$$
$$= \sqrt{n_0 n_1 / (n_0 + n_1)} (\hat{\mu}_1 - \hat{\mu}_0)$$

and

$$\widehat{\text{var}}(\hat{Z}_p) = \frac{n_0 n_1}{n_0 + n_1} \left\{ \widehat{\text{var}}(\hat{\mu}_1) + \widehat{\text{var}}(\hat{\mu}_0) \right\}.$$

Then, the variance of $\hat{\mu}_k$ can be written as

$$\widehat{\text{var}}(\hat{\mu}_k) = \int_0^\tau \left\{ (\tau - s) \frac{1 - \hat{F}_{2k}(s)}{Y_k(s)} - \frac{1}{Y_k(s)} \int_s^\tau \hat{F}_{1k}(u) du \right\}^2 \frac{Y_k(s)}{\hat{S}_k(s)} d\hat{F}_{1k}(s)$$

$$+ \int_0^\tau \left\{ (\tau - s) \frac{\hat{F}_1(s)}{Y_k(s)} - \frac{1}{Y_k(s)} \int_s^\tau \hat{F}_1(u) du \right\}^2 \frac{Y_k(s)}{\hat{S}_k(s)} d\hat{F}_{2k}(s)$$

**Web Appendix 2: Sample size calculation based on RMTLd.**

Let $\mu_k(\tau)$ be the RMTL of the event of interest in group $k$, where $k = 0,1$ correspond to the control group and treatment group, respectively. We define $\text{var}(\mu_k(\tau))$ as the variance of $\mu_k(\tau)$ and $\Delta = \mu_1(\tau) - \mu_0(\tau)$ as the RMTL, and the corresponding variance is $\text{var}(\mu_1(\tau)) + \text{var}(\mu_0(\tau))$. Suppose $n_0$ and $n_1$ are the sample sizes in the control group and treatment group, respectively, and $r = n_1/n_0$ is the ratio of sample sizes. Assume we test the null hypothesis with power $1-\beta$ at a two-sided significance level $\alpha$. Then, under the alternative hypothesis $H_1 : \Delta = \mu_1(\tau) - \mu_0(\tau) \neq 0$, we have

$$
\begin{aligned}
1-\beta &= P\left\{\left|N(\varepsilon,1)\right| > z_{1-\alpha/2}\right\} \approx P\left\{N(\varepsilon,1) > z_{1-\alpha/2}\right\} \\
&= P\left\{N(0,1) > z_{1-\alpha/2} - \varepsilon\right\} \\
&= \Phi(\varepsilon - z_{1-\alpha/2}) \\
&= \Phi\left\{\left[\frac{\mu_1(\tau) - \mu_0(\tau)}{\sqrt{\text{var}(\mu_1(\tau)) + \text{var}(\mu_0(\tau))}}\right] - z_{1-\alpha/2}\right\} \\
&= \Phi\left\{\left[\frac{\mu_1(\tau) - \mu_0(\tau)}{\sqrt{\sigma_1^2 r^{-1}/n_0 + \sigma_0^2/n_0}}\right] - z_{1-\alpha/2}\right\}
\end{aligned}
$$

and $n_0$ can be written as

$$
n_0 = \frac{(z_{1-\beta} + z_{1-\alpha/2})^2}{(\Delta)^2 / (\sigma_0^2 + \sigma_1^2 r^{-1})}.
$$

Hence the total sample size $n$ is $n = n_0 + n_1 = (1+r)n_0$, where $z_p = \Phi^{-1}(p)$ is the inverse standard normal distribution function at probability $p$, and the population variance $\sigma_k^2$ of group $k$ can be estimated as $\hat{\sigma}_k^2 = n_k^* \text{var}^*(\hat{\mu}_k(\tau))$, where $n_k^*$ and $\text{var}^*(\hat{\mu}_k(\tau))$ can be obtained through pilot or previous studies.

**Web Appendix 3**

Inspired by McCaw (2), one can reconstruct the individual subject-level data from Beigel (3) by scanning the Kaplan-Meier curves (Figure 2A in Beigel's paper) of COVID-19 patients using GetData Graph Digitizer software (4) and applying the reconstruction process of Guyot et al. (5).

The reconstructed event times are approximately accurate. The authors reported 399 recoveries on remdesivir and 352 recoveries on the placebo by day 28 (Table S4 in Beigel's paper). The reconstructed data reported 384 recoveries on remdesivir and 333 recoveries on the placebo. Meanwhile, by day 28, the authors reported 59 deaths on remdesivir and 77 deaths on the placebo, respectively, while the reconstructed data reported 62 deaths on remdesivir and 79 deaths on the placebo.

**Web Appendix 4**

**Web Table 1.** The parameter settings of the time to event of interest in simulation design under six scenarios.

| Scenario[a] | Control group | Experimental group |
|---|---|---|
| A | $\exp(1)$ | $\exp(1)$ |
| B | $\exp(1)$ | $\exp(\exp(-0.1))$ |
| C | $\exp(1)$ | $\exp(\exp(-0.3))$ |
| D | $w(1,2)I(t \leq 2) + w(2,2)I(t > 2)$ | $w(4,2)I(t \leq 2) + w(2,2)I(t > 2)$ |
| E | $w(2,2)I(t \leq 2) + w(2,2)I(t > 2)$ | $w(4,2)I(t \leq 2) + w(2,2)I(t > 2)$ |
| F | $w(2,2)I(t \leq 1) + w(2,2)I(t > 1)$ | $w(2,2)I(t \leq 1) + w(0.8,2)I(t > 1)$ |

Abbreviations: exp, exponential distribution; $w$, weibull distribution; $I$, Indicator function.

[a]: A, B, C: Under the proportional subdistribution hazards; D: Early difference; E, F: Late difference

**Web Table 2.** Simulation results of the estimation of RMTLd under proportional SDH

| Sample | CR (%) | A | | | | | B | | | | | C | | | | |
|---|---|---|---|---|---|---|---|---|---|---|---|---|---|---|---|---|
| | | Bias[a] ($\times 10^{-2}$) | Rel bias | RMSE | Rel SE | Cov[b] | Bias[a] ($\times 10^{-2}$) | Rel bias ($\times 10^{-2}$) | RMSE | Rel SE | Cov[b] | Bias[a] ($\times 10^{-2}$) | Rel bias ($\times 10^{-2}$) | RMSE | Rel SE | Cov[b] |
| (300,300) | 0 | -0.0437 | -11.1638 | 0.1289 | 1.0012 | 0.9503 | -0.1116 | 0.2837 | 0.1324 | 0.9887 | 0.9441 | 0.0622 | -0.1210 | 0.1291 | 1.0023 | 0.9509 |
| | 15 | 0.1275 | 32.6121 | 0.1372 | 0.9945 | 0.9487 | 0.0924 | -0.2349 | 0.1378 | 1.0004 | 0.9526 | -0.0159 | 0.0310 | 0.1365 | 0.9959 | 0.9469 |
| | 30 | -0.1657 | -42.3697 | 0.1543 | 0.9819 | 0.9456 | 0.5604 | -1.4239 | 0.1543 | 0.9852 | 0.9471 | 1.3423 | -2.6107 | 0.1502 | 0.9927 | 0.9476 |
| | **45** | **1.2839** | **328.2913** | **0.2178** | **0.9141** | **0.9176** | **5.0395** | **-12.8042** | **0.2176** | **0.9102** | **0.9123** | **9.3131** | **-18.1145** | **0.2227** | **0.9222** | **0.8964** |
| (500,500) | 0 | -0.1462 | -37.3765 | 0.1003 | 0.9977 | 0.9477 | 0.0911 | -0.2314 | 0.1018 | 0.9969 | 0.9469 | 0.0685 | -0.1333 | 0.1020 | 0.9837 | 0.9432 |
| | 15 | 0.1584 | 40.5145 | 0.1049 | 1.0080 | 0.9537 | 0.0974 | -0.2476 | 0.1064 | 1.0044 | 0.9519 | 0.2212 | -0.4303 | 0.1059 | 0.9954 | 0.9482 |
| | 30 | -0.0464 | -11.8561 | 0.1180 | 0.9985 | 0.9503 | 0.7492 | -1.9034 | 0.1196 | 0.9894 | 0.9508 | 1.0726 | -2.0862 | 0.1186 | 0.9763 | 0.9462 |
| | **45** | **0.9078** | **232.1243** | **0.1757** | **0.9027** | **0.9213** | **4.8949** | **-12.4367** | **0.1742** | **0.9195** | **0.9120** | **8.8565** | **-17.2262** | **0.1862** | **0.9044** | **0.8748** |
| (1000,1000) | 0 | 0.0463 | 11.8512 | 0.0718 | 0.9865 | 0.9484 | -0.0598 | 0.1519 | 0.0716 | 1.0029 | 0.9493 | -0.0339 | 0.0659 | 0.0708 | 1.0030 | 0.9540 |
| | 15 | -0.0825 | -21.0924 | 0.0744 | 1.0055 | 0.9499 | 0.2431 | -0.6177 | 0.0746 | 1.0140 | 0.9516 | 0.2058 | -0.4003 | 0.0738 | 1.0108 | 0.9491 |
| | 30 | 0.1286 | 32.8940 | 0.0850 | 0.9846 | 0.9449 | 0.5405 | -1.3734 | 0.0855 | 0.9820 | 0.9459 | 0.8970 | -1.7447 | 0.0827 | 0.9949 | 0.9470 |
| | **45** | **1.1986** | **306.4791** | **0.1262** | **0.9292** | **0.9364** | **5.3642** | **-13.6293** | **0.1345** | **0.9155** | **0.9053** | **9.0107** | **-17.5263** | **0.1486** | **0.9173** | **0.8378** |
| (300,500) | 0 | 0.2184 | 55.8344 | 0.1162 | 0.9943 | 0.9499 | 0.1664 | -0.4228 | 0.1164 | 1.0036 | 0.9501 | 0.2562 | -0.4984 | 0.1159 | 0.9989 | 0.9475 |
| | 15 | -0.0496 | -12.6890 | 0.1229 | 0.9923 | 0.9472 | 0.0852 | -0.2165 | 0.1236 | 0.9952 | 0.9495 | 0.0943 | -0.1835 | 0.1220 | 0.9982 | 0.9490 |
| | 30 | 0.4669 | 119.3765 | 0.1366 | 0.9935 | 0.9477 | 0.6901 | -1.7534 | 0.1380 | 0.9861 | 0.9455 | 1.2735 | -2.4770 | 0.1376 | 0.9750 | 0.9439 |
| | **45** | **1.4044** | **359.0971** | **0.1971** | **0.9139** | **0.9209** | **5.5931** | **-14.2107** | **0.1992** | **0.9151** | **0.9073** | **9.4969** | **-18.4718** | **0.2082** | **0.9193** | **0.8863** |
| (500,1000) | 0 | -0.0903 | -23.0863 | 0.0861 | 1.0072 | 0.9505 | -0.0197 | 0.0500 | 0.0866 | 1.0107 | 0.9503 | -0.1284 | 0.2498 | 0.0872 | 0.9952 | 0.9451 |
| | 15 | -0.0614 | -15.6923 | 0.0904 | 1.0133 | 0.9527 | 0.1646 | -0.4181 | 0.0931 | 0.9914 | 0.9493 | 0.2170 | -0.4221 | 0.0911 | 1.0038 | 0.9506 |
| | 30 | 0.1896 | 48.4697 | 0.1043 | 0.9799 | 0.9467 | 0.6936 | -1.7623 | 0.1029 | 0.9960 | 0.9480 | 1.1150 | -2.1686 | 0.1023 | 0.9897 | 0.9444 |
| | **45** | **1.3641** | **348.8094** | **0.1547** | **0.9040** | **0.9227** | **5.6347** | **-14.3165** | **0.1591** | **0.9151** | **0.9018** | **9.0381** | **-17.5794** | **0.1710** | **0.9120** | **0.8583** |

Abbreviations: Rel bias, relative bias, the mean bias relative to true RMTLd; RMSE, root of the mean squared error; Rel SE, relative standard error, average model standard error/the empirical standard error;, Cov, coverage; CR, censoring rate.

[a]: The true RMTLd under scenario A-C are 0.00004, -0.3935 and -0.5141 years during 4 years, respectively. The relative bias under scenario A appears unstable because the true RMTLd is approximately equal to 0 (in the denominator of relative bias), so we replace bias with relative bias to assess the performance of the RMTLd under scenario A in the text. Meanwhile, when high censoring exists (i.e., 45% censoring), the estimation of the RMTLd is biased and results in undercoverage. The main reason for this phenomenon is that our true RMTLd was chosen to be t = 4 years; that is, the final observed follow-up time was equal to or greater than 4 years in the generation of survival data. Therefore, when high censoring exists, the RMTLd may fail to be estimated because the final follow-up time may be less than 4 years. See associated Web Figure 1 for more details.

[b]: The reasonable coverage is (0.9457, 0.9543) under 10000 simulations.

**Web Table 3.** Simulation results of the estimation of RMTLd under non-proportional SDH

| Sample | CR (%) | D | | | | | E | | | | | F | | | | |
|---|---|---|---|---|---|---|---|---|---|---|---|---|---|---|---|---|
| | | Bias ($\times 10^{-2}$) | Rel bias | RMSE | Rel SE | Cov[b] | Bias ($\times 10^{-2}$) | Rel bias ($\times 10^{-2}$) | RMSE | Rel SE | Cov[b] | Bias ($\times 10^{-2}$) | Rel bias | RMSE | Rel SE | Cov[b] |
| (300,300) | 0 | -0.0599 | 0.2004 | 0.1032 | 1.0206 | 0.9528 | 0.1410 | -0.4009 | 0.1062 | 1.0071 | 0.9523 | 0.0083 | -0.0479 | 0.1085 | 0.9898 | 0.9444 |
| | 15 | -0.1955 | 0.6546 | 0.1120 | 0.9962 | 0.9477 | 0.0972 | -0.2764 | 0.1119 | 0.9967 | 0.9487 | -0.1244 | 0.7193 | 0.1112 | 1.0107 | 0.9512 |
| | 30 | -0.0777 | 0.2601 | 0.1182 | 1.0034 | 0.9512 | 0.1705 | -0.4848 | 0.1170 | 1.0055 | 0.9519 | 0.0810 | -0.4682 | 0.1182 | 1.0061 | 0.9486 |
| | **45** | **1.5385** | **-5.1507** | **0.1361** | **0.9861** | **0.9436** | **0.1329** | **-0.3780** | **0.1291** | **0.9892** | **0.9452** | **-0.1326** | **0.7667** | **0.1317** | **0.9923** | **0.9479** |
| (500,500) | 0 | 0.0640 | -0.2143 | 0.0818 | 0.9985 | 0.9493 | -0.0099 | 0.0281 | 0.0835 | 0.9931 | 0.9485 | -0.0272 | 0.1572 | 0.0836 | 0.9954 | 0.9488 |
| | 15 | -0.1847 | 0.6183 | 0.0861 | 1.0050 | 0.9517 | 0.0971 | -0.2761 | 0.0857 | 1.0089 | 0.9516 | -0.0371 | 0.2146 | 0.0865 | 1.0068 | 0.9513 |
| | 30 | -0.2121 | 0.7100 | 0.0928 | 0.9910 | 0.9471 | 0.0177 | -0.0503 | 0.0912 | 0.9992 | 0.9499 | -0.1120 | 0.6477 | 0.0930 | 0.9912 | 0.9479 |
| | **45** | **1.4651** | **-4.9050** | **0.1072** | **0.9761** | **0.9417** | **-0.0872** | **0.2481** | **0.0987** | **1.0025** | **0.9495** | **-0.1567** | **0.9064** | **0.1012** | **1.0018** | **0.9498** |
| (1000,1000) | 0 | 0.0254 | -0.0850 | 0.0576 | 1.0033 | 0.9511 | 0.0350 | -0.0996 | 0.0591 | 0.9915 | 0.9453 | 0.0046 | -0.0263 | 0.0588 | 1.0011 | 0.9502 |
| | 15 | -0.0081 | 0.0271 | 0.0610 | 1.0038 | 0.9527 | 0.0831 | -0.2364 | 0.0612 | 0.9999 | 0.9526 | -0.0288 | 0.1664 | 0.0617 | 0.9985 | 0.9483 |
| | 30 | -0.1392 | 0.4661 | 0.0655 | 0.9927 | 0.9486 | -0.0752 | 0.2138 | 0.0645 | 1.0000 | 0.9483 | -0.0146 | 0.0847 | 0.0656 | 0.9946 | 0.9493 |
| | **45** | **1.1381** | **-3.8105** | **0.0742** | **1.0041** | **0.9476** | **0.0734** | **-0.2087** | **0.0695** | **1.0075** | **0.9524** | **-0.1873** | **1.0831** | **0.0709** | **1.0115** | **0.9551** |
| (300,500) | 0 | 0.0974 | -0.3262 | 0.0968 | 1.0032 | 0.9490 | 0.0072 | -0.0204 | 0.0953 | 0.9964 | 0.9480 | -0.0623 | 0.3601 | 0.0959 | 0.9937 | 0.9492 |
| | 15 | 0.0201 | -0.0674 | 0.1018 | 1.0084 | 0.9505 | 0.1998 | -0.5680 | 0.0988 | 1.0065 | 0.9521 | -0.0136 | 0.0784 | 0.1001 | 0.9986 | 0.9502 |
| | 30 | 0.0886 | -0.2966 | 0.1096 | 0.9932 | 0.9467 | -0.3235 | 0.9197 | 0.1063 | 0.9905 | 0.9462 | -0.2334 | 1.3494 | 0.1063 | 0.9980 | 0.9498 |
| | **45** | **1.7392** | **-5.8229** | **0.1263** | **0.9817** | **0.9409** | **0.1259** | **-0.3580** | **0.1167** | **0.9842** | **0.9458** | **0.1979** | **-1.1447** | **0.1179** | **0.9919** | **0.9507** |
| (500,1000) | 0 | -0.0201 | 0.0673 | 0.0737 | 0.9992 | 0.9510 | -0.0654 | 0.1859 | 0.0705 | 1.0091 | 0.9498 | 0.0592 | -0.3422 | 0.0712 | 1.0010 | 0.9494 |
| | 15 | -0.1146 | 0.3837 | 0.0791 | 0.9824 | 0.9463 | 0.2077 | -0.5905 | 0.0751 | 0.9924 | 0.9489 | -0.0978 | 0.5657 | 0.0749 | 0.9992 | 0.9490 |
| | 30 | -0.1338 | 0.4481 | 0.0822 | 1.0033 | 0.9487 | -0.0975 | 0.2771 | 0.0787 | 1.0031 | 0.9507 | -0.0832 | 0.4813 | 0.0793 | 1.0029 | 0.9518 |
| | **45** | **1.1887** | **-3.9797** | **0.0948** | **0.9918** | **0.9447** | **-0.0325** | **0.0923** | **0.0859** | **1.0058** | **0.9547** | **-0.2387** | **1.3803** | **0.0892** | **0.9841** | **0.9465** |

Abbreviations: Rel bias, relative bias, the mean bias relative to true RMTLd; RMSE, root of the mean squared error; Rel SE, relative standard error, average model standard error/the empirical standard error; Cov, coverage; CR, censoring rate.

[a]: The true RMTLd under scenario A-C are -0.2986, -0.3517 and -0.1729 years during 4 years, respectively. The relative bias under scenario A appears unstable because the true RMTLd is approximately equal to 0 (in the denominator of relative bias), so we replace bias with relative bias to assess the performance of the RMTLd under scenario A in the text. Meanwhile, when high censoring exists (i.e., 45% censoring), the estimation of the RMTLd is biased and results in undercoverage. The main reason for this phenomenon is that our true RMTLd was chosen to be t = 4 years; that is, the final observed follow-up time was equal to or greater than 4 years in the generation of survival data. Therefore, when high censoring exists, the RMTLd may fail to be estimated because the final follow-up time may be less than 4 years. See associated Web Figure 1 for more details.

[b]: The reasonable coverage is (0.9457, 0.9543) under 10000 simulations.

**Web Figure 1.** Cause-specific hazards (A) and subdistribution hazards (B) of the two groups in example 2. Both the cause-specific hazards (CSHs) and subdistribution hazards (SDHs) of the two groups in example 2 obtained via a nonparametric method (event count/risk set at each event time point). As shown by the hazards curves, both the CSHs and SDHs between groups were non-proportional in the late period; thus, a single constant cHR or sHR may fail to assess the treatment effect with a late difference in this example.

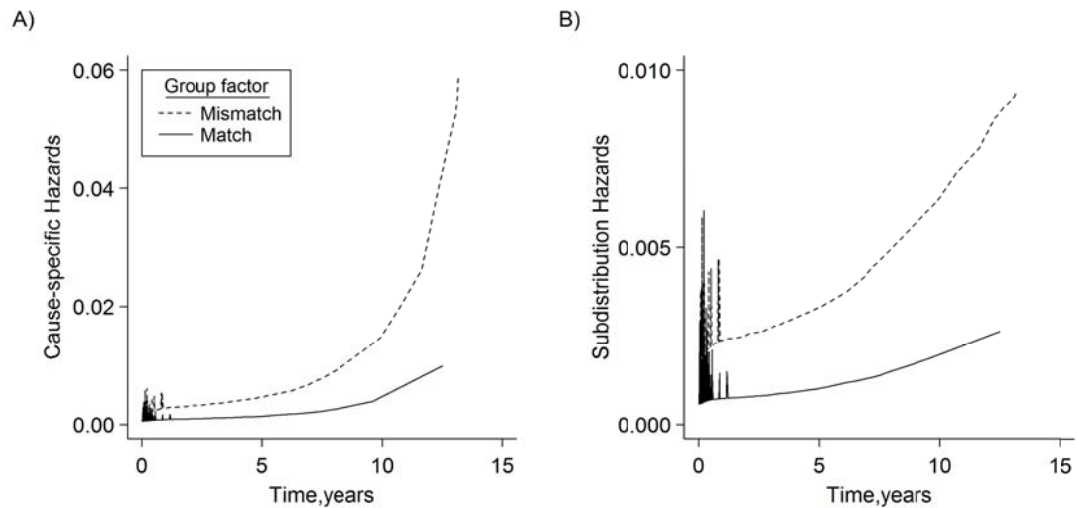

**Web Figure 2.** The cumulative incidence function under different scenarios A - F from simulations. With censoring increases, the final observed follow-up time tends to be less than 4 years. Thus, the RMTLd within $t = 4$ years may fail to be estimated, which results in undercoverage with high censoring (45%).

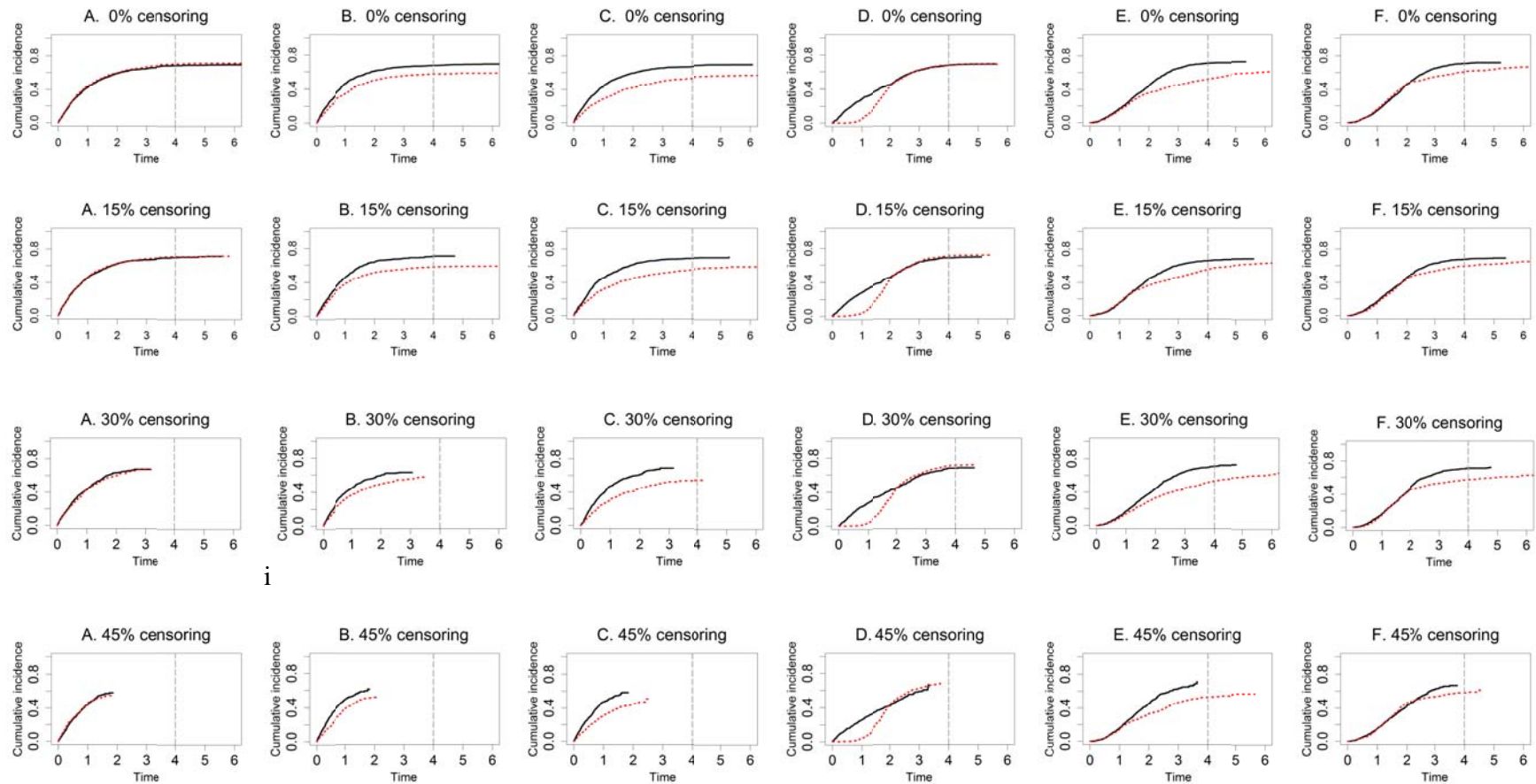

**Web Figure 3.** The sample sizes based on sHR and RMTLd in example 1 and example 2. In Web Figure 3A, the sHR-based and RMTLd-based methods have similar sample sizes when the SDH assumption is satisfied in example 1. In Web Figure 3B, the RMTLd-based sample size decreases as time increases, and after approximately 13.5 years, the RMTLd-based sample sizes are smaller than the sHR-based sample sizes. Combined with the simulation results and Figure 2C, this trend may be interpreted as a late difference in the two CIF curves becoming larger over time, resulting in a smaller sample size over time.

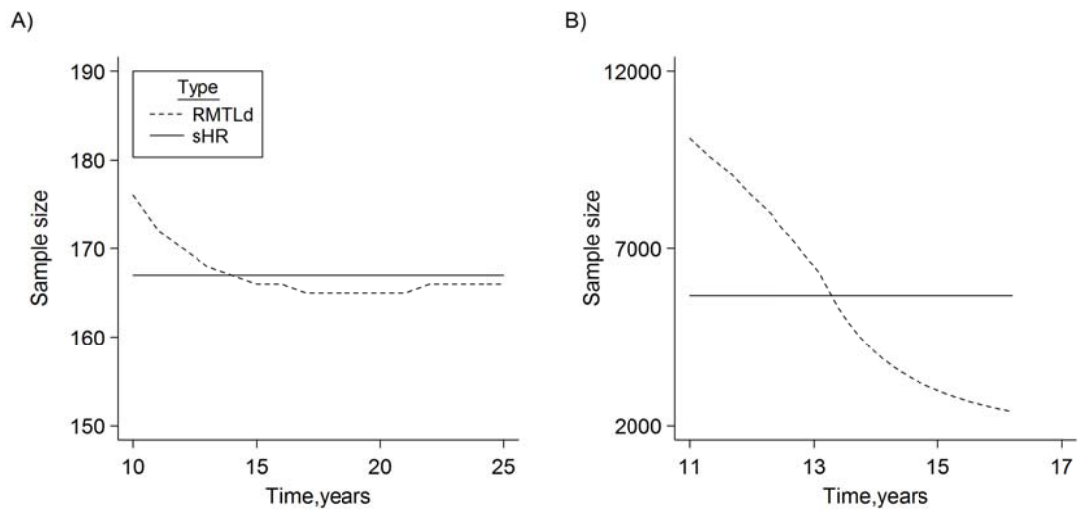

**Reference**

1. Bajorunaite R, Klein J. Two-sample tests of the equality of two cumulative incidence function. Comput Stat Data Anal. 2007;51(9):4269-4281.
2. McCaw ZR, Tian L, Vassy JL, et al. How to Quantify and Interpret Treatment Effects in Comparative Clinical Studies of COVID-19. Ann Intern Med. 2020;173(8):632-637.
3. Beigel JH, Tomashek KM, Dodd LE, et.al. Remdesivir for the Treatment of Covid-19 - Final Report. N Engl J Med. 2020;383(19):1813-1826.
4. GetData Graph Digitizer. http://www.getdata-graph-digitizer.com.
5. Guyot P, Ades AE, Ouwens MJ et al. Enhanced secondary analysis of survival data: reconstructing the data from published Kaplan-Meier survival curves. BMC Med Res Methodol. 2012;12:9.